\documentclass[twocolumn,english,prl,aps,showpacs]{revtex4}
\usepackage[T1]{fontenc}
\usepackage[latin9]{inputenc}
\usepackage{amsmath}
\usepackage{amssymb}
\usepackage{graphicx}
\usepackage{esint}

\makeatletter
\@ifundefined{textcolor}{}
{%
 \definecolor{BLACK}{gray}{0}
 \definecolor{WHITE}{gray}{1}
 \definecolor{RED}{rgb}{1,0,0}
 \definecolor{GREEN}{rgb}{0,1,0}
 \definecolor{BLUE}{rgb}{0,0,1}
 \definecolor{CYAN}{cmyk}{1,0,0,0}
 \definecolor{MAGENTA}{cmyk}{0,1,0,0}
 \definecolor{YELLOW}{cmyk}{0,0,1,0}
 }

\makeatother

\usepackage{babel}
\begin{document}

\title{Quantum Fisher information for density matrices with arbitrary ranks}

\author{Jing Liu, Xiaoxing Jing, Wei Zhong, Xiaoguang Wang}

\address{Zhejiang Institute of Modern Physics, Department of Physics, Zhejiang
University, Hangzhou 310027, China}

\email{xgwang@zimp.zju.edu.cn}

\begin{abstract}
We provide a new expression of the quantum Fisher information(QFI) for a general system. Utilizing this expression, the QFI for a non-full rank density matrix is only determined by its support. This expression
can bring convenience for a infinite dimensional density matrix with a finite support. Besides, a matrix representation of the QFI is also given.

\end{abstract}


\maketitle

\section{Introduction}
Quantum metrology is a field that utilizes the
character of quantum mechanics to improve the precision of a
parameter under detection~\cite{review}. For the past few years,
this field has drawn a lot of attention and has been developing rapidly~\cite{S.loyd,3,5,6,7,8,9,10,11,12,13,14,15,16,17,18}.
Quantum Fisher information(QFI) is a central concept in quantum metrology
because it depicts the lower bound on the variance of the estimator
$\hat{\theta}$ for the parameter $\theta$ due to the Cram\'{e}r-Rao
theorem~\cite{Fisher,FI,FI-1}
\begin{equation}
\mathrm{var}(\hat{\theta})\geq\frac{1}{\nu F},
\end{equation}
where $\mathrm{var(\cdot)}$ is the variance, $\nu$ is the number
of repeated experiments and $F$ is the QFI. However, the QFI is not just limited
in the field of quantum metrology. It has been widely applied in other
aspects of quantum physics~\cite{Luo2004, Luo2006, Paris, Lu, lu2, Ma, Pezze09, wootters, Braunstein,  Pezze, Yu, Toth, mine},
like quantum information and open quantum systems. Thus, it is necessary
and meaningful to study the quantum Fisher information as well as its properties
and dynamical behaviors under various circumstances.

Quantum Fisher information is a local quantity, which can be intuitively interpreted as the  ``velocity" at which the density matrix moves for a given parameter value. This physical interpretation comes from the fact that the QFI is
dependent on the parameterized density matrix $\rho_{\theta}$ and its first derivative $\partial_{\theta}\rho_{\theta}$. Utilizing the spectral decomposition, when the eigenstates of $\rho_{\theta}$ as projectors, act on $\rho_{\theta}$ and its first derivative, the value is only related to the spectral decomposition within the support, which strongly hints that the QFI may be expressed in the representation of the density matrix's support. To find such an expression is the major motivation of this paper.

In this paper, we provide a new expression of the quantum Fisher information
in the representation of the density matrix's support. With this expression,
for a non-full rank density matrix, especially for a infinite dimension one, the QFI may be solved in a finite support space without realizing the knowledge out of the support.
Recently, it is found~\cite{Toth,Yu} that the QFI
can be written in the form of the convex roof of variance. To obtain the QFI,
one should take the minimum value running over all the possible pure-state ensembles.
Utilizing the new expression, we give the condition
when the ensemble from the spectral decomposition is the optimal ensemble in which the minimum value attains. Besides, we also provide a matrix representation form of the QFI and give two examples of it.

\section{Fisher information for a non-full rank density matrix}
In the following we consider a $\mathrm{N}$-dimensional
system ($\mathrm{N}$ can be infinite) with the density operator $\rho_{\theta}$,
which is dependent on the parameter $\theta$. Assume that the spectral decomposition of
the density operator is given by
\begin{equation}
\rho_{\theta}=\sum_{i=1}^{\mathrm{s}}p_{i}|\psi_{i}\rangle\langle\psi_{i}|,\label{spectrum}
\end{equation}
where $p_{i}$ is a eigenvalue and $|\psi_{i}\rangle$ is a eigenstate.
$\mathrm{s}$ is the dimension of the support set of $\rho_{\theta}$, denoted as $\mathrm{supp}(\rho_{\theta})$, i.e., $\mathrm{s}=\mathrm{dim}[\mathrm{supp}(\rho_{\theta})]$.

For a parameterized quantum state $\rho_{\theta}$, the quantum Fisher information
$F$ is defined as below~\cite{FI,FI-1}
\begin{equation}
F:={\rm tr}(\rho_{\theta}L^{2}),\label{eq:df}
\end{equation}
where $L$ is the so-called symmetric logarithmic derivative operator and determined by
\begin{equation}
\partial_{\theta}\rho_{\theta}=\frac{1}{2}\left(L\rho_{\theta}+\rho_{\theta}L\right).\label{eq:rho}
\end{equation}
In the eigenbasis of $\rho_{\theta}$, above equation reads
\begin{equation}
\langle\psi_{i}|\partial_{\theta}\rho_{\theta}|\psi_{j}\rangle
=\frac{1}{2}(p_{i}+p_{j})L_{ij},\label{sld}
\end{equation}
where $L_{ij}:=\langle\psi_{i}|L|\psi_{j}\rangle$. From above equation, one can find that $L_{ij}$ is in principle supported by the full space, but the
value of $L_{ij}$ for $i,j>\mathrm{s}$ is arbitrary because above equation is always established for any value of $L_{ij}$ when $i,j>\mathrm{s}$. Nevertheless, the quantum Fisher information is still a determinate quantity because the calculation of it does not use those values of $L_{ij}$ for $i,j>\mathrm{s}$, which we will show below. Thus, one can set $L_{ij}=0$ for $i,j>\mathrm{s}$ as a matter of convenience.

By substituting Eq.~(\ref{spectrum}) and the normalization relation
$\mathbb{I}=\sum_{j=1}^{\mathrm{N}}|\psi_{j}\rangle\langle\psi_{j}|$ into Eq.~(\ref{eq:df}),
one can obtain the quantum Fisher information as
\begin{equation}
F=\sum_{i=1}^{\mathrm{s}}\sum_{j=1}^{\mathrm{N}}p_{i}L_{ij}L_{ji}.\label{eq:ff}
\end{equation}
Here $\mathbb{I}$ is the identity operator. All $p_{i}$ here is greater
than zero because the index $i\leq \mathrm{s}$ and satisfies $\sum_{i=1}^{\mathrm{s}}p_{i}=1$.
From this equation we see that the randomicity of $L_{ij}$ for $i,j>\mathrm{s}$ does
not affect the certainty of the quantum Fisher information.
As $p_{i}>0$, Eq.~(\ref{sld}) can be rewritten into
\begin{equation}
L_{ij}=\frac{2(\partial_{\theta}\rho_{\theta})_{ij}}{p_{i}+p_{j}},
\end{equation}
where $(\partial_{\theta}\rho_{\theta})_{ij}:=\langle\psi_{i}|\partial_{\theta}\rho_{\theta}|\psi_{j}\rangle$.
Utilizing this expression, Eq.~(\ref{eq:ff}) can be written in the form
\begin{equation}
F_{\theta}=\sum_{i=1}^{\mathrm{s}}\sum_{j=1}^{\mathrm{N}}\frac{4p_{i}}{(p_{i}+p_{j})^{2}}|
(\partial_{\theta}\rho_{\theta})_{ij}|^{2},\label{fisher}
\end{equation}
where the Hermiticity of the operator $\partial_{\theta}\rho_{\theta}$
was used. Next, from the spectral decomposition of $\rho_{\theta}$,
one can find that
\begin{equation}
(\partial_{\theta}\rho_{\theta})_{ij}=\partial_{\theta}p_{i}\delta_{ij}+(p_{j}-p_{i})
\langle\psi_{i}|\partial_{\theta}\psi_{j}\rangle,\label{par}
\end{equation}
where we have used the equation
\begin{equation}
\langle\psi_{i}|\partial_{\theta}\psi_{j}\rangle=-\langle\partial_{\theta}\psi_{i}|\psi_{j}\rangle,\label{eq:syme}
\end{equation}
resulted from the orthogonality $\langle\psi_{i}|\psi_{j}\rangle=\delta_{ij}$. For $i\in [1, \mathrm{s}]$ and $j\in [\mathrm{s}+1, \mathrm{N}]$, the expression of $(\partial_{\theta}\rho_{\theta})_{ij}$ reduces to $-p_{i}\langle\psi_{i}|\partial_{\theta}\psi_{j}\rangle$.
Substituting Eq.~(\ref{par}) into Eq.~(\ref{fisher}), we have
\begin{equation}
F_{\theta} = \sum_{i=1}^{\mathrm{s}}\frac{1}{p_{i}}(\partial_{\theta}p_{i})^{2}+\sum_{i=1}^{\mathrm{s}}\sum_{j=1}^{\mathrm{N}}\frac{4p_{i}(p_{i}-p_{j})^{2}}
{(p_{i}+p_{j})^{2}}|\langle\psi_{i}|\partial_{\theta}\psi_{j}\rangle|^{2}. \label{eq:fisher_c}
\end{equation}

Furthermore, with the knowledge that $\sum_{j=1}^{\mathrm{N}}=\sum_{j=1}^{\mathrm{s}}+\sum_{j=\mathrm{s}+1}^{\mathrm{N}}$,
the second item of above expression can be separated into two parts $F_{1}$
and $F_{2}$. The first part $F_{1}$ reads
\begin{equation}
F_{1}=\sum_{i,j=1}^{\mathrm{s}}\frac{4p_{i}(p_{i}-p_{j})^{2}}{(p_{i}+p_{j})^{2}}|\langle\psi_{i}|\partial_{\theta}\psi_{j}\rangle|^{2},
\end{equation}
 and the second part $F_{2}$ reads
\begin{equation}
F_{2}=\sum_{i=1}^{\mathrm{s}}\sum_{j=\mathrm{s}+1}^{\mathrm{N}}4p_{i}|\langle\psi_{j}|\partial_{\theta}\psi_{i}\rangle|^{2}.
\end{equation}
Based on the normalization relation, it is easy to find that
\begin{equation}
\sum_{j=\mathrm{s}+1}^{\mathrm{N}}|\psi_{j}\rangle\langle\psi_{j}|=\mathbb{I}-\sum_{j=1}^{\mathrm{s}}|\psi_{j}\rangle\langle\psi_{j}|.
\end{equation}
Substituting this equation into the expression of $F_{2}$, one can obtain
\begin{equation}
F_{2}=\sum_{i=1}^{\mathrm{s}}4p_{i}\langle\partial_{\theta}\psi_{i}|\partial_{\theta}\psi_{i}\rangle
-\sum_{i,j=1}^{\mathrm{s}}4p_{i}|\langle\psi_{j}|\partial_{\theta}\psi_{i}\rangle|^{2}.
\end{equation}
Then, the quantum Fisher information can be expressed by
\begin{eqnarray}
F_{\theta} & = & \sum_{i=1}^{\mathrm{s}}\frac{1}{p_{i}}\left(\partial_{\theta} p_{i}\right)^2+
\sum_{i=1}^{\mathrm{s}}4p_{i}\langle\partial_{\theta}\psi_{i}|\partial_{\theta}\psi_{i}\rangle  \notag \\
& & -\sum_{i,j=1}^{\mathrm{s}}\frac{8p_{i}p_{j}}{p_{i}+p_{j}}|\langle\psi_{i}|\partial_{\theta}
\psi_{j}\rangle|^{2}.\quad\label{eq:qFI}
\end{eqnarray}

From this equation one can find that the quantum Fisher information for a non-full rank density matrix is determined by its support. The information of eigenstates
out of the support is not necessary for the calculation of the QFI. This advantage would bring some convenience for the calculation in some cases, especially when $\mathrm{N}$ is infinite and $\mathrm{s}$ is finite.

According the theory of the classical Fisher information~\cite{Fisher,FI,FI-1}, it is natural to treat the first item of Eq.~(\ref{eq:qFI}) as the classical contribution of quantum Fisher information~\cite{Paris} because $\sum_{i=1}^{\mathrm{s}}\frac{1}{p_{i}}(\partial_{\theta}p_{i})^{2}=4\sum_{i=1}^{\mathrm{s}}
\left(\partial_{\theta}\sqrt{p_{i}}\right)^{2}$.
Then the quantum Fisher information for a quantum system can be  separated into
two parts, the classical contribution and quantum contribution, namely,
\begin{equation}
F_{\theta}=F_{\mathrm{ct}}+F_{\mathrm{qt}},
\end{equation}
where the classical contribution reads
\begin{equation}
F_{\mathrm{ct}}=\sum_{i=1}^{\mathrm{s}}4\left(\partial_{\theta}\sqrt{p_{i}}\right)^{2},\label{eq:cFI}
\end{equation}
and the quantum contribution reads
\begin{equation}
F_{\mathrm{qt}}=\sum_{i=1}^{\mathrm{s}}4p_{i}\langle\partial_{\theta}\psi_{i}|\partial_{\theta}\psi_{i}\rangle
-\sum_{i,j=1}^{\mathrm{s}}\frac{8p_{i}p_{j}}{p_{i}+p_{j}}|\langle\psi_{i}|\partial_{\theta}
\psi_{j}\rangle|^{2}. \label{eq:q}
\end{equation}

The separation of the quantum Fisher information is not just in form.
From the equations above, one can find that the classical contribution
of the quantum Fisher information is a special case of the classical
Fisher information. It can be treated as the classical Fisher information
obtained through the measurement $\{|\psi_{i}\rangle\}$ in the eigenspace
of $\rho_{\theta}$: $\mathbb{E}^{\mathrm{N}}$. The eigenspace $\mathbb{E}^{\mathrm{N}}$ is
spanned by the basis $\left\{|\psi_{i}\rangle \right\}$, and
$\left\{p_{i}\right\}$ is a classical distribution in this space.
From Eq.~(\ref{eq:cFI}), it is not difficult to find that the classical contribution $F_{\mathrm{ct}}$ is only related to the derivative of the eigenvalues, which indicates
that this part of information is coming from the classical distribution
in $\mathbb{E}^{\mathrm{N}}$. Moreover, we find that the classical contribution has the following properties:
(1) it vanishes for pure states;
(2) it vanishes for the unitary parametrization;
(3) it is invariant under unitary transformation of density matrix,
no matter the transformation is parameter-dependent or not.

In the mean time, with some transformation, Eq.~(\ref{eq:q}) can be rewritten as
\begin{equation}
F_{\mathrm{qt}}=\sum_{i=1}^{\mathrm{s}}p_{i}F_{Q}(|\psi_{i}\rangle)-\sum_{i\neq j}^{\mathrm{s}}\frac{8p_{i}p_{j}}{p_{i}+p_{j}}|\langle\psi_{i}|\partial_{\theta}
\psi_{j}\rangle|^{2}, \label{eq:fqt}
\end{equation}
where
\begin{equation}
F_{Q}(|\psi_{i}\rangle)=4\left(\langle\partial_{\theta}\psi_{i}|\partial_{\theta}\psi_{i}\rangle
-|\langle\psi_{i}|\partial_{\theta}\psi_{i}\rangle|^{2}\right)
\end{equation}
is the quantum Fisher information of the eigenstate $|\psi_{i}\rangle$.
From this equation, it is clear that $F_{\mathrm{qt}}$ is related to the basis of $\mathbb{E}^{\mathrm{N}}$. In $\mathbb{E}^{\mathrm{N}}$, $F_{\mathrm{qt}}$ is determined by the weighted average of all the quantum Fisher information $F_{Q}(|\psi_{i}\rangle)$ of the basis vector $|\psi_{i}\rangle$ and the coupling between these vectors. This manifests that this part of information originates from the quantum structure of space $\mathbb{E}^{\mathrm{N}}$. These are the geometric meanings of the classical and quantum contribution as well as the intrinsic reason for the separation.

We know the classical contribution of the QFI always vanishes for the unitary parametrization. But for a non-unitary parametrization procedure, including the channel estimation~\cite{channel_es,channel_es1,channel_es2,channel_es3,channel_es4,channel_es5} and the noise estimation~\cite{Chaves,NJP}, the classical contribution does have an influence on the precision. However, only improving the classical contribution without enhancing the quantum counterpart, the precision is not available to surpass the shot-noise limit, the lower limit for a total classical scenario. The estimation of the decoherence strength~\cite{NJP}, in which the classical contribution plays the leading role, is an example of this scenario.

The quantum Fisher information is a local quantity, which can be intuitively interpreted as the ``velocity" at which the matrix moves for a given parameter value. In mathematics, this means that the quantum Fisher information depends on the density matrix $\rho_{\theta}$ and its first derivative $\partial_{\theta}\rho_{\theta}$. Utilizing the spectral decomposition, there exists items such as $|\psi_{i}\rangle\langle\partial_{\theta}\psi_{j}|$ and $|\partial_{\theta}\psi_{i}\rangle\langle\psi_{j}|$.
When these items are traced with the eigenstates out of the support, the values turn out to be zero.  This is the intuitive reason that the QFI can be expressed in the representation of the support. If the QFI is related to the higher order derivatives, like the second one  $\partial^2_{\theta}\rho_{\theta}$, then there would exist the item like $|\partial_{\theta}\psi_{i}\rangle\langle\partial_{\theta}\psi_{j}|$. As $|\partial_{\theta}\psi_{i}\rangle$ is not always orthogonal with $|\psi_{j}\rangle$, when this item is traced with the projectors out of the support, the value cannot always be zero, then the quantum Fisher information has to be related to the whole Hilbert space, rather than the support only.

For the unitary parametrization $\exp(i\theta H)$, the classical contribution vanishes, and the quantum Fisher information reduces to
\begin{equation}
F_{Q}=\sum_{i=1}^{\mathrm{s}}p_{i}F_{Q}(|\psi_{i}\rangle)-\sum_{i\neq j}^{\mathrm{s}}\frac{8p_{i}p_{j}}{p_{i}+p_{j}}|\langle\psi_{i}|H|
\psi_{j}\rangle|^{2} \label{eq:fqt_1}.
\end{equation}
In the mean time,  $F_{Q}(|\psi_{i}\rangle)$ reduces to the form that is proportional to the variance of operator $H$ on the eigenstates, i.e.,
\begin{equation}
F_{Q}(|\psi_{i}\rangle)=4(\Delta H)^2_{|\psi_{i}\rangle},
\end{equation}
where $(\Delta H)^2_{|\psi_{i}\rangle}:=\langle\psi_{i}|H^2|\psi_{i}\rangle
-|\langle\psi_{i}|H|\psi_{i}\rangle|^2$ is the variance.
Recently, T\'{o}th and Petz~\cite{Toth} found that for a rank-2 system the quantum Fisher information can be treated as the convex roof of the variance, then Yu~\cite{Yu} proves that
this theorem is also established for a general system, namely,
\begin{equation}
F_{\theta}=\min_{\{q_{k}, |\Psi_{k}\rangle\}}4\sum_{k}q_{k}(\Delta H)^2_{|\Psi_{k}\rangle} \label{eq:Toth_eq}.
\end{equation}
Here $\{q_{k}, |\Psi_{k}\rangle\}$ refers to a set of pure-state ensembles, which satisfies
\begin{equation}
\rho_{\theta}=\sum_{k}q_{k}|\Psi_{k}\rangle\langle\Psi_{k}|.
\end{equation}
One should notice that the ensemble of the eigenvalues and eigenstates $\{p_{i}, |\psi_{i}\rangle\}$ is one of these ensembles,
but not the only one. Comparing Eq.~(\ref{eq:fqt_1}) with Eq.~(\ref{eq:Toth_eq}), one can find that the condition for the ensemble $\{p_{i}, |\psi_{i}\rangle\}$ being the optimal ensemble is that the transition item
\begin{equation}
\langle\psi_{i}|H|\psi_{j}\rangle=0, \mbox{ for any } i\neq j.
\end{equation}
For example, in some Mach-Zehnder interferometer, $H=\frac{1}{2i}(a^{\dagger}b-ab^{\dagger})$, where $a$, $b$ are the annihilation operators of two modes, and $a^{\dagger}$, $b^{\dagger}$ are the creation operators respectively. Choosing an appropriate input state, like an even state~\cite{mine} or a Fock state~\cite{Pezze} in one port, the item $\langle\psi_{i}|H|\psi_{j}\rangle$ vanishes for any $i\neq j$, then the ensemble $\{p_{i}, |\psi_{i}\rangle\}$ is the optimal ensemble and the QFI reduces to $F_{\theta}=4\sum_{i=1}^{\mathrm{s}}p_{i}(\Delta H)^2_{|\psi_{i}\rangle}$.

This condition can be checked through another way. Based on Ref.~\cite{Yu}, we introduce an observable
\begin{equation}
\mathrm{Y} = \sum_{i,j}\frac{2\sqrt{p_{i}p_{j}}}{p_{i}+p_{j}}H_{ij}|\psi_{i}\rangle\langle\psi_{j}|,
\end{equation}
where $H_{ij}=\langle\psi_{i}|H|\psi_{j}\rangle$. Denote the spectral decomposition $\mathrm{Y}=\sum_{k}\alpha_{k}|y_{k}\rangle\langle y_{k}|$, then the optimal pure state can be constructed as
\begin{equation}
|U_{k}\rangle = \frac{1}{\sqrt{u_{k}}}\sum_{i}U_{ki}\sqrt{p_{i}}|\psi_{i}\rangle,
\end{equation}
with $u_{k}=\sum_{i}|U_{ki}|^2 p_{i}$ and $U_{ki}=\langle\psi_{i}|y_{k}\rangle$. When $|U_{k}\rangle=|\psi_{k}\rangle$, there must be $|y_{k}\rangle=|\psi_{k}\rangle$. As $|y_{k}\rangle$ is the eigenstate of observable Y, then one can see that the condition for $|y_{k}\rangle=|\psi_{k}\rangle$ is that all the off-diagonal elements of observable Y have to vanish, i.e., $H_{ij}=0$ for any $i\neq j$, which coincides with our result.

\section{Matrix representation}
In this section we show a matrix representation
of the quantum Fisher information. We consider the classical contribution first. Define a $\mathrm{N}$-dimensional diagonal matrix $D$ with elements $D_{ii}=p_{i}$, then the classical contribution can be rewritten in the form
\begin{equation}
F_{\mathrm{ct}}=4\mathrm{Tr}\left(\partial_{\theta}\sqrt{D}\right)^{2}.
\end{equation}
This equation is equivalent to Eq.~(\ref{eq:cFI}) as $p_{i}=0$ for $i\in [\mathrm{s}+1,\mathrm{N}]$.

Define a $\mathrm{N}$-dimensional
matrix $\mathcal{P}$ with the elements $\mathcal{P}_{ij}:=|\langle\psi_{i}|\partial_{\theta}\psi_{j}\rangle|^{2}$.
It is easy to see that the matrix $\mathcal{P}$ is real and symmetric.
The symmetry can be proved by using Eq.~(\ref{eq:syme}) into the
definition above.
Denote a constant $\mathrm{N}$-dimensional matrix $\mathcal{I}$ whose elements
are 1, i.e., $\mathcal{I}_{ij}=1$ for any $i$ and $j$, and define
a $\mathrm{N}$-dimensional block diagonal matrix $\mathcal{G}$, which is
$\mathcal{G}=\mathrm{diag}[\mathcal{H}_{\mathrm{s}\times \mathrm{s}},0_{(\mathrm{N}-\mathrm{s})\times(\mathrm{N}-\mathrm{s})}]$,
where $\mathcal{H}_{\mathrm{s}\times \mathrm{s}}$ is a $\mathrm{s}$-dimensional real symmetric
matrix. The elements of $\mathcal{H}$ are the harmonic mean values,
$\mathcal{H}_{ij}=2p_{i}p_{j}/(p_{i}+p_{j}).$
With the help of above matrices, as well as the symmetry of $\mathcal{P},$
i.e., $\mathcal{P}_{ij}=\mathcal{P}_{ji}$, the quantum contribution
can be written in the form
\begin{equation}
F_{\mathrm{qt}}=4\mathrm{Tr}\left[\left(D\mathcal{I}-\mathcal{G}\right)\mathcal{P}\right].\label{eq:quamtum}
\end{equation}
This is the matrix representation of quantum contribution of the QFI.
It is easy to see that the coefficient matrix $D\mathcal{I}-\mathcal{G}$ is traceless.

The matrix $\mathcal{P}$ can be treated as the ``transfer'' matrix
between the vector of the eigenstates $(|\psi_{1}\rangle,\cdots,|\psi_{i}\rangle,\cdots,|\psi_{\mathrm{N}}\rangle)^{\mathrm{T}}$
and its derivative vector. For a unitary parametrization,
the element of $\mathcal{P}$ reads $\mathcal{P}_{ij}=|\langle\phi_{i}|H|\phi_{j}\rangle|^{2}$.
In this case, the diagonal element of $\mathcal{P}$ is the survive
probability of the eigenstate $|\phi_{i}\rangle$ under the evolution
$H$ and the non-diagonal element is the transition probability between
$|\phi_{i}\rangle$ and $|\phi_{j}\rangle$ under $H$.

Compared with Eqs.~(\ref{eq:cFI}) and (\ref{eq:q}), the matrix
representation of the quantum Fisher information is related
to the entire $\mathrm{N}$-dimensional space. However, the coefficient matrix
$D$, $\mathcal{G}$ and the ``transfer'' matrix $\mathcal{P}$
are all real and symmetric. For a unitary parametrization, in the matrix representation,
one does not need to calculate the average value of $H^{2}$, but the transition item
$\langle\psi_{i}|H|\psi_{j}\rangle$ has to be calculated through the entire space, not
only those in the support. In the mean time, using the expression of Eq.~(\ref{eq:q}),
one has to calculate the average value of $H^{2}$ under the eigenstates, but the transition item
needn't to be calculated out of the support. These two representations have their own merits and will bring convenience if being used properly.

In the following we give two examples
utilizing this matrix representation. First we apply it in the qubit case. In this case, the parameterized density matrix $\rho_{\theta}$ can be decomposed
as $\rho_{\theta}=\sum_{i=1}^{2}p_{i}(\theta)|\psi_{i}(\theta)\rangle\langle\psi_{i}(\theta)|$.
Then the coefficient matrix reads
\begin{equation}
D\mathcal{I}-\mathcal{G}=\left(\begin{array}{cc}
0 & p_{1}-2\det\rho_{\theta}\\
p_{2}-2\det\rho_{\theta} & 0
\end{array}\right),
\end{equation}
where the equation $p_{1}p_{2}=\det\rho_{\theta}$ has been used.
Thus, it is easy to obtain the quantum contribution as
\begin{equation}
F_{\mathrm{qt}}=4\left(1-4\det\rho_{\theta}\right)\mathcal{P}_{12},\label{eq:qFI-qubit}
\end{equation}
where $\mathcal{P}_{12}=|\langle\psi_{1}|\partial_{\theta}\psi_{2}\rangle|^{2}$.

When the state is a pure state, for instance $p_{1}=1$ and $p_{2}=0$,
there is $\det\rho_{\theta}=0$, then the quantum contribution reduces to
\begin{equation}
F_{\mathrm{qt}} = 4\mathcal{P}_{12}=4|\langle\psi_{1}|\partial_{\theta}\psi_{2}\rangle|^{2}.
\end{equation}
This form coincides with the traditional quantum Fisher
information form for pure state: $F_{Q}=\langle\partial_{\theta}\psi_{1}|\partial_{\theta}\psi_{1}\rangle
-|\langle\psi_{1}|\partial_{\theta}\psi_{1}\rangle|^{2}$,
which can be proved by substituting the normalization relation $\mathbb{I}
=|\psi_{1}\rangle\langle\psi_{1}|+|\psi_{2}\rangle\langle\psi_{2}|$
into the item $\langle\partial_{\theta}\psi_{1}|\partial_{\theta}\psi_{1}\rangle$.
The classical contribution can also be obtained in this case,
which reads
\begin{equation}
F_{\mathrm{ct}}=\frac{\left(\partial_{\theta}p_{1,2}\right)^{2}}{\det\rho_{\theta}}
=\frac{\det\rho_{\theta}}{1-4\det\rho_{\theta}}
\left[\partial_{\theta}\left(\ln\det\rho_{\theta}\right)\right]^2  \label{eq:cFI-qubit}
\end{equation}
for mixed states and $F_{\mathrm{ct}}=0$ for pure states. For a unitary parametrization, the quantum contribution reads
\begin{equation}
F_{\mathrm{qt}}=4\left(1-4\det\rho_{0}\right)|\langle\phi_{1}|H|\phi_{2}\rangle|^{2},
\end{equation}
with $|\phi_{i}\rangle$ a eigenstate of $\rho_{0}$. As $D$ is
independent of $\theta$, the classical contribution vanishes for
both mixed and pure states.

Next we give another example. Consider a density matrix with the following form~\cite{Hyllus}
\begin{equation}
\rho_{\theta}=\sum_{n=0}^{\infty}Q_{n}\rho_{\theta}^{(n)},
\end{equation}
where $Q_{n}$ is a real number and independent of $\theta$. $\rho_{\theta}^{(n)}$
is a state of $n$ particles in the entire Hilbert space. This form
is representative for an optical system taking into account the superselection
rules~\cite{Hyllus}. For a unitary parametrization,
the spectral decomposition of $\rho_{\theta}$ reads
\begin{equation}
\rho_{\theta}=\sum_{n=0}^{\infty}\sum_{i=0}^{n}Q_{n}q_{i}^{(n)}|\psi_{i}^{(n)}\rangle\langle\psi_{i}^{(n)}|,
\end{equation}
where $|\psi_{i}^{(n)}\rangle=e^{-iH\theta}|\phi_{i}^{(n)}\rangle$.
In this case, the classical contribution vanishes. If the transition
between the eigenstates in different particle spaces through the Hamiltonian $H$
is forbidden, which is feasible in some cases~\cite{Jzrzyna}, namely, $\langle\phi_{i}^{(n)}|H|\phi_{j}^{(n^{\prime})}\rangle=0$
when $n\neq n^{\prime}$, then the ``transfer'' matrix $\mathcal{P}$
can be written in a block diagonal form $\mathcal{P}=\sum_{n=0}^{\infty}\mathcal{P}^{(n)}$,
where $\mathcal{P}^{(n)}$ is the corresponding ``transfer'' matrix
for fixed $n$ particles. According to the feature
of trace operation, only the corresponding block diagonal part of
the coefficient matrices $D$, $\mathcal{I}$ and $\mathcal{G}$ matters in the
calculation of the quantum contribution.
If we define $D^{(n)}$, $\mathcal{I}^{(n)}$ and $\mathcal{G}^{(n)}$
as the coefficient matrices for fixed $n$ particles, then the block
diagonal parts of $D$, $\mathcal{I}$
and $\mathcal{G}$ can be expressed as $\sum_{n}Q_{n}D^{(n)}$, $\sum_{n}\mathcal{I}^{(n)}$
and $\sum_{n}Q_{n}\mathcal{G}^{(n)}$. Thus, the quantum Fisher information
reads
\begin{equation}
F_{Q}=4\sum_{n=0}^{\infty}Q_{n}\mathrm{Tr}\left[\left(D^{(n)}\mathcal{I}^{(n)}-\mathcal{G}^{(n)}\right)\mathcal{P}^{(n)}\right].
\end{equation}
Also, one can find that the quantum Fisher information $F^{(n)}$ in
the subspace of fixed $n$ particles can be written as
\begin{equation}
F_{Q}^{(n)}=4\mathrm{Tr}\left[\left(D^{(n)}\mathcal{I}^{(n)}-\mathcal{G}^{(n)}\right)\mathcal{P}^{(n)}\right].
\end{equation}
Thus, one can write the total quantum Fisher information in the form
\begin{equation}
F_{Q}=\sum_{n=0}^{\infty}Q_{n}F_{Q}^{(n)}.
\end{equation}
This total quantum Fisher information is the weighted average of all the quantum Fisher
information for fixed $n$ particles. This form of the QFI has been widely used in the optical interferometry devices when no external global phase reference is present~\cite{Demkowicz09}.

More generally, taking into account the transition between the eigenstates in different
particle subspaces, $\mathcal{P}$ can still be separated into blocks according to
the particle number. Denote the sub-block in the upper and lower triangular
of $\mathcal{P}$ between $n$ and $n^{\prime}$ particle subspaces as
$\mathcal{P}^{(nn^{\prime})}$ and $\mathcal{P}^{(n^{\prime}n)}$, respectively.
The diagonal block $\mathcal{P}^{(n)}$ is the same as above.
Then, $\mathcal{P}$ can be expressed in the form
$\mathcal{P}=\sum_{n}\mathcal{P}^{(n)}+\sum_{n\neq n^{\prime}}\mathcal{P}^{(nn^{\prime})}$,
so as $\mathcal{I}$ and $\mathcal{G}$. Here
all the elements of $\mathcal{P}^{(nn^{\prime})}$ is non-negative
based on the property of $\mathcal{P}$. Thus, the total quantum Fisher information
can be written as
\begin{equation}
F_{Q}=\sum_{n}Q_{n}F_{Q}^{(n)}+\sum_{n\neq n^{\prime}}4\mathrm{Tr}\left[C^{(n n^{\prime})}\mathcal{P}^{(n^{\prime}n)}\right],\label{eq:temp}
\end{equation}
where $C^{(nn^{\prime})}=Q_{n}D^{(n)}\mathcal{I}^{(nn^{\prime})}-\mathcal{G}^{(nn^{\prime})}$.

From this equation one can find that when all the elements of $C^{(nn^{\prime})}$ is non-negative, the transition between the eigenstates in different particle subspaces, i.e., the second item of Eq.~(\ref{eq:temp}), can enhance the total QFI. Apart from this condition, the effect has to be discussed case by case.

\section{Conclusion}
In this paper, we provide a new analytic expression of the quantum Fisher information. For a non-full rank density matrix, this new expression is only determined by the support of the density matrix. With this new expression, the QFI for some infinite systems can be solved in a finite support space. This would bring significant advantage during the calculation in some scenarios. Besides, we also provide a matrix representation form of the quantum Fisher information and give two examples.

\begin{acknowledgments}
This work was supported by NFRPC through Grant
No. 2012CB921602, the NSFC through Grants No. 11025527 and No. 10935010.

\end{acknowledgments}

\end{document}